# Quantum mechanics and modelling of physical reality.

**Marian Kupczynski***

Département de l'Informatique, UQO, Case postale 1250, succursale Hull, Gatineau. QC, Canada J8X 3X7
* Correspondence: marian.kupczynski@uqo.ca;

**Abstract:** Quantum mechanics led to spectacular technological developments, discovery of new constituents of matter and new materials: however there is still no consensus on its interpretation and limitations. Some scientists and scientific writers promote some exotic interpretations and evoke quantum magic. In this paper we point out that magical explanations mean the end of the science. Magical explanations are misleading and counterproductive. We explain how a simple probabilistic locally causal model is able to reproduce quantum correlations in Bell tests. We also discuss difficulties of mathematical modelling of the physical reality and dangers of incorrect mental images. We examine in detail when and how a probabilistic model may describe completely a random experiment. We give some arguments in favor of contextual statistical interpretation of quantum mechanics. We conclude that we still don't know whether the quantum theory provides a complete description of physical phenomena and we explain how it may be tested. We also point out that there remain several open questions and challenges which we discuss in some detail. In particular there is still no consensus about how to reconcile quantum theory with general relativity and cosmology.

**Keywords:** Statistical interpretation of quantum mechanics and problem of completeness,, EPR-Bohm paradox, Bell-type inequalities and quantum non-locality, locally causal explanation of quantum correlations, testing predictable completeness of quantum mechanics, problems with a description of physical reality: abstract models and incorrect mental images, quantum magic, successes and challenges of quantum theory, semi-empirical models and free parameters, infinities in quantum field theory and renormalization procedures.

1. Introduction

Quantum mechanics (QM) is extremely successful theory. Since it uses abstract mathematical language several physicists, philosophers and scientific writers add to it interpretations which are sometimes pure science fiction and general public is highly confused. On Wikipedia one may find more than 18 interpretations of QM. In our opinion many of these interpretations add only to the confusion.

We adopt minimalistic statistical contextual interpretation (SCI) which combines in some sense statistical interpretations of Einstein [1, 2], and Ballentine [3, 4] with Copenhagen interpretation (CI) promoted by Bohr [5]. In this interpretation [6-14] a quantum state is not an attribute of a single physical system which can be changed instantaneously but it describes statistical properties of an ensemble of similarly prepared systems. Quantum states, which are mathematical entities, together with Hermitian operators representing measurements of various physical observables, in well-defined experimental contexts, allow making objective probabilistic predictions on a statistical scatter of measurement outcomes. SCI and CI do not provide any detailed space-time description of quantum phenomena.

In SCI and in CI a wave function of the universe is simply a non-sense. According to SCI and CI it is a non-sense to say that an electron may be <u>at the same time</u> here and a meter away or to compare entangled pair of photons to two perfectly random dice that always give the correct matching outcomes.

All quantum paradoxes are based on incorrect images of quantum phenomena and/or on incorrect interpretation of a quantum state vector (wave function). SCI agrees with Einstein [1] : …Ψ function does not, in any sense, describe the state of one single physical system.." and with Ballantine [4]: "…the habit of considering an individual particle to have its own wave function is hard to break.... though it has been demonstrated strictly incorrect". Reduced wave functions describe different sub-ensembles of physical systems

In SCI it is justified to ask whether more detailed locally causal description of quantum phenomena is possible. According to SCI one may not even take for granted that QM is *predictably complete* (complete from a predictability point of view). Therefore one should carefully check whether quantum probabilities grasp all reproducible properties of experimental time-series.

John Bell [15, 16] demonstrated that some probabilistic local realistic hidden variables probabilistic models (LRHV) are unable to reproduce all quantum predictions for spin polarization correlation experiments (SPCE). Namely he proved some inequalities for spin correlations which could not be violated by the correlations deduced using LRHV but which should be violated, for some experimental setting, by the correlations predicted by QM. Several other Bell-type inequalities were proven and tested in carefully designed experiments.

Experiments [16-21] confirmed the predictions of QM and the violation of inequalities. Several physicists concluded that no rational causally local explanation of these long range quantum correlations might be given. Therefore it seems that either correlations are created due to the quantum magic and come from out of the space-time or we are living in a correlated super-deterministic universe and the experimentalists' freedom to choose their experimental settings is an illusion.

Nevertheless rational explanation of violation of Bell- type inequalities has been given since many years ago, often independently, by several authors [6-11, 22-60]. Several computer simulations were able to reproduce quantum predictions [61-68]. A recent review may be found in {33, 40, 41, 51-52}. In spite of this a belief in in the existence of nonlocal instantaneous influences has not been fading and it nourishes the hope that a scalable superfast quantum computer using these influences may be constructed.

There is nothing magical in the violation of Bell -type inequalities. The probabilistic models used to prove inequalities are not general enough and they are inconsistent with the experimental protocols used in SPCE [51]. If supplementary parameters describing measuring instruments are correctly incorporated in local probabilistic models than inequalities may not be proven [38, 51, 52, 56, 57] . Similarly if one does not assume that all experimental outcomes are predetermined and measuring instruments register passively these outcomes then all non-probabilistic proofs of these inequalities are also not valid [8].

One may find in serious scientific papers statements that Nature is nonlocal, that when three quantum particles are put in two boxes yet no two particles are in the same box etc. Quantum magic is promoted in several blogs, videos on YouTube, in BBC awarded documentary entitled: "Where that could be identical copies of you?" etc. One might say that magic sells better!

The aim of this paper is to point out that if we abandon searching for rational explanation of physical phenomena and if we evoke magic then the science will become a science fiction.

Therefore people working on the quantum computer project or trying to reconcile general relativity with QM have to be aware that rational explanations of quantum correlations do exist. They should also know that the Bohr-Einstein quantum debate about the completeness of quantum mechanics may not be closed [52].

We know now that QM does not give a complete description of physical systems since we need quantum electrodynamics, quantum field theory and the standard model. The Bohr-Einstein quantum debate is about how we are modelling physical reality and in what sense a statistical description of phenomena may be considered complete

Einstein agreed that it is impossible to measure simultaneously a position and a linear momentum of an electron with an arbitrary precision but believed that a complete theory should not abandon a space-time description of its invisible motion. Bohr pointed out that space-time lost its empirical foundation at the atomic scale and only abstract mathematical description provided by QM may be used to explain quantum phenomena. .

In 2002 we wrote a paper on the completeness of quantum mechanics [46]. Some parts of this paper were extended and published later. It still contains some unpublished material which we include after some editing in this article. It discusses the difficulties of mathematical modelling of physical reality and it gives some insight how one may understand and test the completeness of statistical description of phenomena. In particular we point out that we do not even know whether QM is predictably complete and that it should be tested.

We are not alone to be preoccupied with the present situation. Let us quote here for example Schlosshauer, Kofler and Zeilinger [69] who analyzed the results of a poll carried out in 2013 among 33 participants of a conference on the foundations of quantum mechanics: "Quantum theory is based on a clear mathematical apparatus, has enormous significance for the natural sciences, enjoys phenomenal predictive success, and plays a critical role in modern technological developments. Yet, nearly 90 years after the theory's development, there is still no consensus in the scientific community regarding the interpretation of the theory's foundational building blocks. Our poll is an urgent reminder of this peculiar situation."

Already Richard Feynman said: "Nobody understands quantum mechanics". If we continue to use incorrect mental images and imprecise terminology we will never do.

The paper is structured as follows.

In Section 2 we talk about modelling of the physical reality and about the origin of Bell inequalities.

In Section 3 we explain on an example of a simple locally causal probabilistic model why Bell type inequalities may not be proven and QM predictions may be reproduced

In Section 4 we discuss what it means that a probabilistic model gives a complete description of a random experiment and how it may be tested.

Section 5 contains a discussion of some open questions and challenges.

## 2. Modelling of the physical reality.

In this section we reproduce after some editing several paragraphs form [46].

Let us imagine that we are sitting on a shore of an island on a lake watching a sunset. We see bird's flying, leaves and branches are moving with a wind, a passage of a boat produces all interesting patterns on the surface of the lake and we hear regular waves hitting the shore. Finally a big round circle of the sun is hiding under the horizon leaving a place for beautifully illuminated clouds and later for planets and stars. All these physical phenomena are perceived by us in three dimensions and they are changing in time usually in an irreversible way.

To do physics we have to construct mathematical models leading to predictions concerning our observations and measurements. This is why we created concepts of material points, waves and fields. For Newton light was a stream of small particles. For Maxwell light was an electromagnetic wave moving in a continuous invisible medium called ether, similarly to waves on a water. With the abandon of ether in special theory of relativity we lost the intuitive image of the propagation of light.

A discovery of the fact that exchanges of energy and of linear momentum between light and matter are quantized gave a temptation to represent again light as a stream of indivisible (point-like) photons moving rectilinearly and being deflected only on material obstacles or absorbed and emitted by atoms. This picture together with the assumption that each indivisible photon may pass only by one slit or another and that the interaction with a slit through which it is passing <u>does not depend on the fact that other slit is open or closed</u> is clearly inconsistent with the observed interference pattern. Photons are not localizable objects but similar interference phenomena may be observed with electrons, with C-60 molecules [70] etc.

We discovered that the light and the matter may present wave or corpuscular behavior in mutually exclusive (complementary) experimental arrangements. It is clear that we cannot model a light source as a gun and photons as bullets. Similarly if we observe a passage of a boat on a lake we can detect and even measure the energy and the momentum transferred by regular waves hitting a buoy close to a shore. We could even think that we observe a beam of "wavelons" hitting the buoy close to a shore. However there is no comparable transfer of energy and momentum on any buoy, floating in deep water, away from a shore. Therefore we cannot make an image of the boat producing a beam of "wavelons". Of course we can see changes on the surface of the lake but in quantum physics we do not see the" lake". This example shows a danger of image making. Wrong images lead to paradoxes and to wrong deductions.

There is wholeness in quantum experiments. The only picture given by QM is a black box picture. As an input we have an initial "beam" of physical systems entering a box as an output we have a modified" beam" ("beams") or a set of counts on various detectors. QM does not give any intuitive detailed picture of what is happening in the box. Let us cite Bohr [5]:"Strictly speaking, the mathematical formalism of quantum mechanics and electrodynamics merely offers rules of calculation for the deduction of expectations pertaining to observations obtained under well-defined experimental conditions specified by classical physical concepts".

This statement is valid not only for the description of many standard atomic phenomena but also for S- matrix description of all scattering processes of elementary particles and for stochastic models describing the time evolution of trapped molecules, atoms or ions. The quantum mechanics and new stochastic approaches have no deterministic prediction for a single measurement or for a single time -series of events observed for a trapped ultra-cold atom. The predictions being of statistical or of stochastic character apply to the statistical distribution of results obtained in long runs or in several repetitions of the experiment.

A classical mechanics also concentrates on the quantitative description of observations.. The Sun and the Earth are represented mathematically by material points characterized at each moment of time by their masses, positions and velocities. If in some inertial frame initial positions and velocities are known Newton's equations allow determining a subsequent motion of these points which agrees remarkably well with the real motion of the Earth around the Sun. <u>There is no speculation by what mechanism a change in the position of one body causes an instantaneous change in the acceleration of another body</u> but it does not harm the success of the model. Of course a quest for a more detailed understanding of the mutual interactions between distant masses led to the progress in physics namely to the development of classical electrodynamics and to the creation of the general theory of relativity.

In spite of the fact that QM gives often only statistical predictions on outcomes of various experiments a claim is made that QM gives a complete description of the physical phenomena and even the most complete description of individual physical systems. Einstein has never accepted this claim and his famous paper written with Rosen and Podolsky [71] started the discussions on the interpretation and completeness of QM. These discussions continue till now.

In the statistical interpretation [4], inspired by Einstein, a wave function describes only an ensemble of identically prepared physical systems and the reduction of wave function is a passage from the description of the whole ensemble of these systems to the description of a sub-ensemble satisfying some additional conditions. The statistical interpretation is free of paradoxes because a single measurement does not produce the instantaneous reduction of the wave function. The statistical interpretation leaves a place for the introduction of supplementary parameters (called often hidden variables) which would determine the behavior of each individual physical system during the experiment. Several theories with supplementary parameters (TSP) have been discussed [15].

QM gives predictions for spin polarization correlation experiments (SPCE) dealing with pairs of electrons or photons produced in a singlet state. In order to explain these long range correlations Bell analyzed a large family of TSP so called local or realistic hidden variable theories (LRHV) and showed that their predictions must violate quantum mechanical predictions for some configurations of the experimental set-up. Bell's argument was tested in several experiments and confirmed the predictions of quantum mechanics.

Many physicists conclude that if a TSP wants to explain experimental data it must allow for faster than light influences between particles and thus violate Einstein's locality. Even without deep reasoning one can see that this conclusion must be flawed. Let us imagine a huge volcanic eruption taking place somewhere in the middle of the Pacific Ocean. Tsunami waves hitting the shores of Japan and America will be correlated in a natural way. Long range correlations usually are due to a common cause and come from the memory of past events and time evolution. They do not require superluminal influences between distant objects.

It was shown by many authors that the assumptions made in LRHV are more restrictive, that they seemed to be, and that Bell's inequalities may be violated not only by quantum experiments but also by macroscopic ones. Here we stop this s edited long citation from [46].

Since 2002 the violation of Bell-type inequalities has been confirmed in several experiments [18-21] and has been explained in a rational way in several books and articles. Nevertheless, as we mentioned in the introduction many authors still incorrectly believe that one has to abandon the locality of Nature or experimenters' freedom of choice. In a recent paper it was clearly explained why such beliefs are unfounded [52]. In the next section we recall a simple causally local contextual probabilistic model for twin photon beam SPCE which is able to reproduce predictions of QM and does not allow proving any Bell -type inequality.

**3. A local and causal model explaining quantum correlations**

In twin photon beam SPCE experiments a source is emitting two correlated signals arriving to distant polarization beam splitters (PBS) and detectors. The signals produce clicks on the detectors and the correlations between the clicks registered by Alice and Bob are estimated for different settings of their PBS.

In spite of what was claimed one may construct a simple locally causal probabilistic model reproducing these correlations [52]. It must include some additional parameters (contextual hidden variables) which describe signals, PBS and detectors at the moment of their interaction. Clicks or their absence for a given setting (x, y) are determined in a causal and local way in function of correlated parameters ($\lambda_1, \lambda_2$) describing signals and uncorrelated parameters ($\lambda_x, \lambda_y$) describing measuring devices at the moment of a ``measurement''. We assume also that we are not living in super-deterministic universe and that experimentalists may choose experimental settings (x, y) as they wish: randomly or in a systematic way. In our model it does not matter. The choice of (x, y) does not depend on ($\lambda_1$, $\lambda_2$) describing incoming signals.

Simple probabilistic model incorporating contextual hidden variables allows reproducing the long range correlations between clicks observed by Alice and Bob in different settings [52, 53]:

$$E(A,B | x, y) = \sum_{\lambda \in \Lambda_{xy}} A_x(\lambda_1, \lambda_x) B_y(\lambda_2, \lambda_y) P_x(\lambda_x) P_y(\lambda_y) P(\lambda_1, \lambda_2) \quad (1)$$

where $\Lambda_{xy} = \Lambda_1 \times \Lambda_2 \times \Lambda_x \times \Lambda_y$, $A_x(\lambda_1, \lambda_x)$ and $B_y(\lambda_2, \lambda_y)$ are equal $0, \pm 1$. Random experiments performed in different settings are described using different parameter spaces $\Lambda_{xy}$ in agreement with QM and with Kolmogorov theory of probability [10, 34, 45, 51]. The oversimplification made by Bell was the assumption that $\Lambda_{xy} = \Lambda = \Lambda_1 \times \Lambda_2$ what is equivalent to assuming that the correlations in different incompatible experiments may be deduced from a joint probability distribution on some unique probability space $\Lambda$ [25, 26]. If the model (1) is used Bell-type inequalities may not be proven and quantum predictions may be reproduced. In spite of the fact that it has been explained by several authors the speculations about nonlocality of Nature and the quantum magic continue.

Using the model (1) one obtains explicitly local expectation values of single clicks observed by Alice

$$E(A | x) = \sum_{\lambda_1, \lambda_x, \lambda_2} A_x(\lambda_1, \lambda_x) P_x(\lambda_x) P(\lambda_1, \lambda_2) \quad (2)$$

and Bob:

$$E(B | y) = \sum_{\lambda_1, \lambda_y, \lambda_2} B_y(\lambda_2, \lambda_y) P_y(\lambda_y) P(\lambda_1, \lambda_2) \quad (3)$$

The model (1-3) allows explaining why single counts deduced from the estimated correlations between Alice's and Bob's may depend on the settings and it does not mean that Einsteinian no-signaling is violated [53].

Several authors argue, evoking the Bayes theorem, that setting dependent parameters compromise experimenters' freedom of choice. It has been proven recently in [52] that such conclusion is incorrect.

Therefore the violation of Bell-type inequalities neither proves the nonlocality of Nature nor the completeness of QM.

In the next section we discuss a general problem whether a probabilistic description of a random experiments may be considered complete. We conclude that we still don't know whether QM is predictably complete and that it should be tested.

## 4. Can a statistical description of phenomena be considered complete?

In this section we reproduce again after some editing several paragraphs form [46].

A statistical description is not a description of individual objects but it is a description of regularities observed in large populations or in the outcomes of a series of repeated random experiments.

Let us examine a series of coin flipping experiments. Instead of coins having head and tails we have coins with one side "blue"(B) and one side "red"(R). If we want to provide a complete description of a coin using concepts of classical physics and mechanics we may say that a coin is a round disk of a given diameter. We can find also its mass, volume, moment of inertia etc. All these attributes (values of classical observables) describe "completely" a coin from a classical point of view. We have also at our disposal various flipping devices. From outside all of them look the same: you have a place to put a coin, one of the faces up, and a button to push on. A coin is projected and you see it flying, rotating and finally it lands on an observation plate.

- EXPERIMENT 1 (E1). We start with a device D1 and we use only one coin. At first we do not pay attention what is a color of a face of the coin which we put up. For example we record a series of outcomes: BBRBRRRB... At the first sight it is a time series of events without any regularity. We decide now to be more systematic and to put always a face B up into the device. To our big satisfaction we obtain a simple time series: RRRRRR... If instead we put a face R up into the device we obtain: BBBBB.... From an empirical point of view our description of the phenomenon is complete. A device D1 is a classical deterministic device such that if we insert into it a particular coin it changes a face B up into a face R up and vice versa. However we do not see only the final result we see also a coin flying, revolving and landing. If you are a physicist you would like to understand why so complicated phenomenon gives a simple deterministic result. Let us imagine that we are allowed to examine the interior of the device. If we see that D1 gives always to the coin the same initial linear velocity and the same initial angular velocity then knowing the laws of classical mechanics and taking into consideration air resistance but neglecting the influence of the air turbulences, caused by a revolving coin, we can, with a help of a computer, reproduce a flight of the coin and deduce that the coin placed with one face up will land always on the observation plate with another face up. It would provide a complete description of the phenomenon. Even if we were unable to make calculations we could anticipate a result and we could say that we understood " completely" the studied phenomenon. Of course we took the Newton's equations for granted but in some point looking for the explanation we have to stop asking a question: "Why?"

- EXPERIMENT 2 (E2). We take the same coin and a device D2. On a basis of the previous experiment we start by placing the coin always with face B up and we preform several series of trials. To our surprise we get a time series of results BRBRBRB...or RBRBR...We obtain similar results if we place the face R up. A complete empirical explanation of the phenomenon is that D2 produces completely deterministic alternating series of outcomes. The only uncertainty is a first result. It shows that a device has some memory. For example a flipping mechanism of the D2 can be identical to the flipping mechanism of the D1 with one difference that the inserted coins are rotated around a horizontal axis before being flipped with a rotating mechanism keeping a memory of events: each 180◦ rotation is followed by 360◦ and vice versa. To understand ''completely" the phenomenon we examine the interior of the device and we repeat the analysis we did for E1.

- EXPERIMENT 3 (E3). We replace the device D2 by a device D3. We repeat several times the experiment with the face B up and after with the face R up. We obtain various time series which seem to be completely random. We call a colleague statistician for help. He checks that the observed time series is random. He observes that relative empirical frequencies of observing the face B in long runs are close to 0.5. It concludes that each experiment is a Bernoulli trial with a probability p=0.5. Using this assumption he can make predictions concerning the number of faces B observed in N-repetitions of the experiment and compare them with the data. A statistical description of the observed time-series of results is complete and it may be resumed in the following rigorous way: Anytime we place the coin into the device D3 there are two outcomes possible each obtained with a probability 0.5. A probability 0.5 it is not the information about the coin. It is not the information about the device D3. It is only the information about the statistical distribution of outcomes of random experiments: inserting the coin into the device, pushing the button on and registering the outcome. This is why a statement: the coin, if flipped, has a probability 0.5 to land with the face B up is incorrect. All devices considered above are flipping devices but the statistical distributions of the results they produce are different. We could correct this statement by adding: if flipped with the device D3, but one has to remember what it means. Once again to understand completely the phenomenon we could look in the interior of the device D3 .For example we might find that D3 is identical to the device D1 but before flipping there was some mechanism rotating the coin in a pseudo random way. It would allow us to understand" more completely" the phenomenon but it would not give us any additional information about the statistical distribution of results. There could be however an advantage of this "more complete" description of the phenomenon

Let us imagine that to each device considered above we add a ventilator blowing on the coin when it is flying. It would certainly modify statistical distribution of results in the experiments E1 and E2. From the empirical point of view the device D1 with a ventilator it is a new device D'1 so we have a new random experiment and a new statistical distribution to be found. However on this level we are unable to predict how this new description originates from a previous one. On the contrary a knowledge of the "complete " description of the phenomenon describing a flight of the coin produced by D1 could be used to predict the modifications induced by the wind produced by the ventilator. If we had a classical theory describing time evolution of the air turbulences and its interactions with the coin (which we don't have) we could in principle determine possible trajectories of the coin and deduce the changes in the statistical distribution of experimental outcomes.

In all these experiments we saw the coin flying and we could look inside the experimental devices. If we did not have this knowledge but only the knowledge of final results the only unambiguous description would be a statistical one. Probably we could invent infinite number of "microscopic" hidden variable models agreeing with observations but we would not gain any better understanding of the results.

This resembles the situation in quantum mechanics. We have a stable source producing some beam. We place in front of a beam some detector which clicks regularly what makes us believe that we have a beam of some invisible "particles" having some constant intensity. We take the detector out and we pass our beam by the experimental arrangement (a device) and we observe a time– series of the possible final outcomes. QM gives us algorithms to calculate probability distributions of outcomes giving no information how a time-series is building up. Einstein understood very well the statistical description of the experiments given by QM but he believed that this statistical description should be "completed" by some "microscopic" description explaining how the observed time- series of the results is building up.

It seems to us that if such description existed, it would be extremely complicated and not unique so perhaps not very useful. Even if one does not think that such "microscopic " description is needed a hypothesis, that such description is possible, suggests that there is some information in the time- series of the results not accounted for by the statistical description given by QM. If it was true a careful analysis of time- series of outcomes could reveal some structure not explained by QM what would imply that statistical description provides an incomplete statistical description of the data.

Therefore a question whether a particular statistical description of the phenomenon is complete or not, it is an experimental question which can be asked and answered independently of the existence of a "microscopic" description of the phenomenon. The answer can be obtained with the help of the purity tests which were proposed many years ago [42, 43] and never done. We explain the idea of these tests continuing our discussion of simple experiments with coins.

As we saw in the experiment E3 any time- series obtained could be interpreted as a series of results of consecutive repetitions of identical Bernoulli trials each characterized completely by a probability p = 0.5. Let us consider now another random experiment.

- EXPERIMENT 4 (E4). There is a box containing coins but we do not see what is in the box. There are 51 blue coins and 51 red coins in the box .With closed eyes we mix well coins, we draw one coin from the box, we place it on a table and finally we open eyes and we record the color of a coin without replacing it in the box. We continue drawing the coins and when 100 coins are on the table we return all of them to the box. If we repeat this random experiment several times we find that ,each time, a frequency of drawing a blue coin is close to 0.5. We are tempted to conclude that a probability of drawing a blue coin in each draw is equal to 0.5. The probabilistic description of the experiments E4 and E3 seems to be identical. Our friend statistician tells us not to jump into conclusion too fast because if initially in the box we have 2N coins (N red and N blue) on the average we find 50% of blue coins in a sample but a time- series is different and in principle we can discover it by a more detailed statistical analysis of this series. In the case of Bernoulli trials at each repetition the probability of drawing a blue coin is the same. On the contrary in E4 the probability of drawing a blue coin in the k-th draw depends on a number of blue coins drawn already. If among $k$ draws there were $m$ blue coins then the probability of obtaining a blue coin in the next draw is p(k+1)=p(k+1,m, N)= (N-m)/(2N-k). <u>Thus in E4 we have a succession of different dependent random experiments when in the E3 we have a succession of identical independent random experiments.</u> The averages of two time series are consistent but the time- series are different. In this case a statement that a probability of drawing a blue face is in each draw in E4 is equal to 0.5 is not only incomplete but it is also incorrect.

If we modify the experiment E4 namely by returning a coin to the box after each draw our new experiment is, for samples of a size smaller than 102, completely equivalent to the experiment E3. On the "microscopic " level there is however one fundamental difference: in E4 the coins in the box are always either blue or red when the coin in the experiment E3 is neither blue nor red but unfortunately this difference is not seen from the existing data. To see more easily how such "microscopic" differences could be detected by performing additional experiments we discuss another macroscopic experiments with coins.

- EXPERIMENT 5 (E5) There is a box, which contains now 50 blue coins and 50 red coins having all other physical properties identical. A button is pushed and a mechanical arm picks at random one of the coins in a box and inserts it into the device D3. The result B or R is recorded and handed to the experimenter and a coin is returned to the box.

- EXPERIMENT 6 (E6) The only difference with E5 is that instead of containing 50 red and 50 blue coins a box contains 100 two-sided coins identical to the coin used in the experiments E1-E3. All other physical properties of two-sided coins are the same as the physical properties of the coins in E5.

The experiments E5 and E6 produce finite time-series of results which do not allow to find any significant difference between them. Two physicists agree with this statement but they cannot agree how to interpret the results. One of them, a follower of Einstein, says: we have a statistical mixture (mixed statistical ensemble) in the box of the same number of blue and red coins and because we draw the coins from the box with replacement thus on average we observe 50% of blue coins in each run of the experiment.

A second physicist, a follower of Bohr, says: it is nonsense we have simply a pure statistical ensembles of quantum coins each in the same pure quantum state, such that each of them has simply a probability 0.5 to become blue or red after interacting with the measuring device. They meet a statistician who confirms that the experiments give indistinguishable results and tell them that without performing supplementary experiments one cannot decide whose model is a correct one.

He tells them that <u>in a mixed statistical ensemble some of it sub-ensembles can in principle have different observable statistical properties</u>. On the contrary if one has a pure statistical ensemble all of its sub-ensembles have the same properties as the initial ensemble. Our physicists agree with the statistician and they make a hole in the boxes containing coins and they decide to remove the same number of coins in E5 and in E6 before proceeding with several repetitions of their experiment. If they removed by chance the equal number of blue and red coins in E5 no difference could be noticed but if by chance they changed a proportion of blue coins in their box then they could see a difference in long runs of the experiment. If fewer coins were left in the box differences could be bigger. For example with 4 blue and 6 red coins in the box the probability of outcome B is 0.4 instead of 0.5. With one blue coin in the box they would get p=1. Following the same protocol for the experiment E6 they would not register any significant difference in the results. After performing these experiments they may confidently conclude that there is a "microscopic" difference between E5 and E6. Namely Einstein's model applies to E5 and Bohr's model may apply to E6.

Therefore an unproven claim that QM gives a complete description of an individual quantum system may not be disproved by any philosophical argument nor by a mathematical theorem but it may only be disproved by experimental data. A probabilistic description of experiments performed on an ensemble of identically prepared individual systems can be said to describe completely an interaction of an individual system with the experimental device when a statistical ensemble is pure and a time series of data is a simple random sample.

The assumption of completeness of the statistical description provided by QM is not only unnecessary but it is counter-productive. The experimentalists are interested only in testing the statistical distributions of experimental results in long runs without even trying to analyze in detail observed time-series. They eliminate "bad" experimental runs, sometimes without finding any logical reason for doing it, simply because in the theory there is no place for them. We have enormous amount of data accumulated. If we performed tests of the randomness and the purity tests [43] on these data perhaps we would discover new statistical regularities in the time- series we had never thought they existed. Here we stop this edited long citation from [46].

We are now in 2018 and still the homogeneity of experimental random samples has not been studied carefully enough. In the meantime new tests which could be used to check whether QM is predictably complete were explained [72-74]. We also demonstrated with Hans de Raedt [75] that sample inhomogeneity leads to a dramatic breakdown of the standard statistical analysis and makes statistical significance tests inconclusive. This is why we hope that testing sample homogeneity will become one day an essential part of any analysis of experimental data.

The violation of Bell inequalities confirmed a contextual character of quantum observables. Quantum phenomena and the measurement outcomes are produced when physical systems interact with various instruments in well-defined experimental contexts.

QM discovered the existence of incompatible physical observables which are represented by non -commuting operators. In order to measure these observables one has to use mutually exclusive experimental set-ups and the variances of the statistical distributions of the measurement outcomes are related by uncertainty relations. An important couple of such observables are the position and the linear momentum another couple is spin-projections in two different directions.

There is no experimental set-up allowing to measure simultaneously and with arbitrary precision the values of these incompatible physical observables on any physical system. A joint probability distribution of random variables representing incompatible physical observables does not exist thus it cannot be used to reproduce all the predictions of QM for mutually exclusive experiments in which the values of these variables are measured. It was confirmed by the violation of Bell-type inequalities and by the negativity of the Wigner function.

There is a big difference between a position of an electron and its spin projection. A projection of electron's spin takes only discrete values and is the effect of the interaction of electron's magnetic moment with the Stern-Gerlach apparatus [52]. An electron is often considered to be a "point-like particle" thus by its definition at any moment of time it must be somewhere. This unknown position can never be measured with an arbitrary precision and we find only a macroscopic ionization track in the bubble chamber or in the emulsion caused by an electron's passage. Electron is not only a mass and a charge but it is surrounded by the electromagnetic field. Therefore one should not forget that the configuration of this field depends on how many slits are open in the two slit experiment before saying that an electron is passing by two distant slits at the same time.

According to Copenhagen interpretation a question by which slit electron is passing through is meaningless but it does not mean that an electron can be in two distant positions at the same time and it appears only in a well localized position if we make a position measurement asking: Electron where are you?" This is why we believe as, Einstein did, that the Moon is there even if we do not look at it.

Particle physicists consider electrons, protons and heavy ions as a "real stuff" and prepare various beams of them in accelerators (operating according to the laws of classical electrodynamics) to study their collisions. They use mathematical algorithms provided by quantum electrodynamics and by the Standard Model to "explain" their observations in a quantitative way. <u>It is important to point out that Feynman graphs are only simple pictorial tools which allow including all contributions to perturbative calculation of transition amplitudes.</u> By no means are they <u>faithful</u> images of real processes taking place in a space- time.

Similarly trajectories in the path integral formulation of quantum mechanics are only mathematical entities which may hardly justify the faith in so called many world interpretation of QM which starts to be fashionable again.

**4. Discussion**

We got used to think that our perceptions are some imperfect images of the underlying objective physical reality. Mathematical models of classical physics used mathematical entities which were idealizations of our observations and provided the description of the physical reality which was consistent with our common sense and with the local causally we have been witnessing in our everyday experience.

To explain invisible world of atoms and elementary particles we succeeded to create quantum mechanics, quantum electrodynamics and quantum field theory (QFT) which allowed us to provide a quantitative description of many physical phenomena and to predict the existence of new phenomena. Quantum theories use complicated mathematical models and often give only probabilistic predictions on a statistical scatter of experimental outcomes. Mathematical models do not contain intuitive images and explanations how observed phenomena and individual experimental outcomes, registered by macroscopic instruments, are produced.

Various scientists add to the abstract quantum description contradictory interpretations and explanations. One of the challenges is to arrive to the consensus which of these interpretations is correct [69, 76]. Some scientists claim that we have simply to accept quantum magic.

We believe that there is no person who was not amazed by tricks performed by a professional magician. We do not understand how he is doing it but we know that there is a causally local explanation of these apparent miracles. Similarly Einstein strongly believed there should be some locally causal explanation of quantum miracles.

In this paper we recalled a locally causal explanation of one of these miracles: nonlocal correlations between entangled quantum systems. Einstein objected a claim that a statistical description of physical phenomena may be considered complete. In this paper we discussed in some detail how such claim may be tested and we pointed out that in fact we don't even know whether QM is predictably complete [43, 72].

There is another serious problem. The mathematics is a rigorous theory but often exact solutions of mathematical equations cannot be found. This problem we encountered when we tried to solve Newton's equations of motion, Schrodinger equations, interacting quantum field equations etc. Thus we had to develop techniques to obtain approximate solutions.

QFT is unable to describe exactly the scattering of bound states. Therefore semi- empirical models containing several adjustable parameters are added to a theory in order to explain various phenomena in particle physics. In particular the comparison of the Standard Model with experimental data is a difficult task requiring many free parameters, various phenomenological inputs and Monte Carlo simulation of events [77, 78].

All models for high-energy baryon-baryon and heavy ion collisions assume that the Optical Theorem is valid [79]. It was clearly demonstrated that one may have a unitary S-matrix without the Optical Theorem thus the validity of the Optical Theorem should not be taken for granted and should be carefully tested [79-85]. The violation of the Optical Theorem might be easily reconciled with the standard model [85].

Standard Model faces also serious challenges related to the discovery of black matter, massive neutrinos, teraquarks and pentaquarks.

There is also another problem: quantum models are very flexible and allow introduction of several free parameters in order to explain experimental data. Therefore there is a danger that quantum theory becomes unfalsifiable [86].

Quantum theory led to spectacular technological developments, discovery of new constituents of matter and new materials and we may be proud of it. Quantum-like probabilistic models find successful applications in psychology, economy and in other domains of science [11, 41].

However we should not forget what Bohr said :"The main point to realize is that a knowledge presents itself within a conceptual framework adapted to account for previous experience and that any such frame may prove too narrow to comprehend new experiences"[5].

Moreover in QFT we encounter several infinities which are removed by various well understood renormalization procedures. The infinities arrive because the fields are defined on a continuous space-time and we are dealing with point-like charges and masses. It would be much more elegant to construct a theory which does not require any renormalization. This was the opinion of Dirac who at the end of his book wrote:" the difficulties being of a profound character can be removed only by some drastic change in the foundations of the theory, probably a change as drastic as the passage from Bohr's orbit theory to the present quantum mechanics" [4, 87].

Perhaps such drastic change will be needed to reconcile the quantum theory with the general relativity and cosmology.

We mentioned only few open questions and challenges standing before quantum theory. When this paper was completed we found that exhaustive discussion of various open questions was also given recently by Allen and Lidstrom [88], by Coley [89] and by Khrennikov [90].

The correct understanding of the foundations of quantum theory is also important for the success of the quantum computer program.